\documentclass[preprint,12pt]{elsarticle}

\usepackage{graphicx}
\usepackage{dcolumn}
\usepackage{bm}
\usepackage{amssymb}

\journal{Physics Letters B}

\newcommand*{\no}{\noindent}
\newcommand*{\bea}{\begin{eqnarray}}
\newcommand*{\eea}{\end{eqnarray}}
\newcommand*{\be}{\begin{equation}}
\newcommand*{\ee}{\end{equation}}
\newcommand*{\pd}{\partial}
\newcommand*{\pdm}{\pd_{\mu}}

\newcommand*{\pref}[1]{(\ref{#1})}

\newcommand*{\mn}{{\mu\nu}}

\newcommand*{\tr}{\mathrm{tr}}

\begin{document}
\begin{frontmatter}

\title{Constructing non-perturbative gauges using correlation functions}

\author{Axel Maas}
\address{Department of Theoretical Physics, Institute of Physics,\\
             Karl-Franzens University Graz, Universit\"atsplatz 5, A-8010 Graz,
             Austria}

\begin{abstract}
Gauge fixing in the non-perturbative domain of non-Abelian gauge theories is obstructed by the Gribov-Singer ambiguity. To compare results from different methods it is necessary to resolve this ambiguity explicitly. Such a resolution is proposed using conditions on correlation functions for a family of non-perturbative Landau gauges. As a consequence, the various results available for correlation functions could possibly correspond to different non-perturbative Landau gauges, discriminated by an additional non-perturbative gauge parameter. The proposal, the necessary assumptions, and evidence from lattice gauge theory calculations, are presented in detail.
\end{abstract} 


\end{frontmatter}

\section{Introduction}

To describe gauge-dependent degrees of freedom, like quarks and gluons, it is necessary to fix a gauge. In the non-perturbative domain of non-Abelian gauge theories this is complicated. Local gauge conditions, e.\ g.\ for Landau gauge
\be
\pd^\mu A_\mu^a=0,\label{landau}
\ee
\no are satisfied by more than one field configuration $A_\mu^a$, being Gribov copies of each other \cite{Gribov:1977wm}. Conditions, which are not local, are required to resolve this ambiguity \cite{Singer:dk}.

It is desirable that such supplemental conditions are formulated independent of the particular algorithmic implementation of a method. Furthermore, in the context of lattice calculations and functional methods, a formulation using correlation functions would be desirable. In the perturbative case the latter is possible. E.\ g., \pref{landau} can be formulated as the vanishing of the longitudinal part of the gluon propagator $D_\mn$,
\be
p_\mu p_\nu D^\mn=0.
\ee
\no Also, certain supplemental conditions for the Landau gauge can be formulated in this form. One example is the absolute Landau gauge, which requires to absolutely minimize the total trace of the gluon propagator \cite{Maas:2008ri,vanBaal:1997gu}
\be
\tr D=c\int d^dp D_{\mu}^\mu(p),\label{trd}
\ee
\no with $c$ a positive constant. However, \pref{trd} requires regularization in the continuum and it is not clear whether this introduces new ambiguities. Thus, alternatives are desirable. This should in general be possible by conditions on correlation functions: Given a set (of configurations) of sets of distinct events (Gribov copies identified by the gauge field at every space-time point), it is always possible to construct probability distributions (gauges selecting a particular Gribov copy) such that they are distinguished and identified uniquely by at least some of their - finite or infinite - moments (correlation functions). The remaining question is whether this possibility can be cast into a practical method, in particular whether the distinguishing correlation functions can be determined.

In continuum studies a one-parameter family of correlation functions has been obtained \cite{Boucaud:2008ky,Fischer:2008uz}, distinguished by the ghost propagator, an infinite-order moment with respect to the elementary gluon field. This motivates the present study whether this family could indeed arise as the consequence of a gauge choice, as speculated in \cite{Maas:2008ri,Fischer:2008uz}. Indeed, at least in a finite volume a corresponding family is constructed using gauge conditions resolving, at least partly, the Gribov-Singe ambiguity. Assuming the (qualitative) correctness of the results in the continuum this would suggest a one-to-one correspondence.

\section{The structure of the residual gauge orbit}

After imposing the Landau gauge condition \pref{landau} remains a set, called the residual gauge orbit \cite{Maas:2008ri} here, of Gribov copies separated by large gauge transformations. These orbits can be divided in Gribov regions \cite{Gribov:1977wm}, with the orbit of each configuration passing at least once through each region \cite{Dell'Antonio:1991xt}. A possible first step to construct a definite gauge-fixing prescription is to restrict the residual orbit to one of these regions. Here, the first Gribov region where the Euclidean Faddeev-Popov operator $-\pdm D_\mu$, with the covariant derivative $D_\mu$, is positive semi-definite is chosen.

For this restriction an explicit prescription purely in terms of Green's functions is not yet known. However, general considerations for a positive operator and all results so far (for a recent compilation of references see \cite{Maas:2008ri,Fischer:2008uz}) strongly suggest that it is sufficient that the ghost propagator $D_G$, being the expectation value of the inverse Faddeev-Popov operator, is required to be of definite (negative) sign and possibly monotonous. An explicit evaluation in 1+1-dimensional Coulomb gauge \cite{Reinhardt:2008ij} is finding exactly that the ghost propagator changes sign for some momenta when evaluated outside the first Gribov region. However, lattice calculations by construction \cite{Cucchieri:2006tf} and functional methods implicitly \cite{Zwanziger:2003cf} ensure that results are obtained from inside the first region, and this question is of minor concern. Hence, it is not necessary to obtain the spectrum of the Faddeev-Popov operator explicitly.

\begin{figure}
\includegraphics[width=0.5\linewidth]{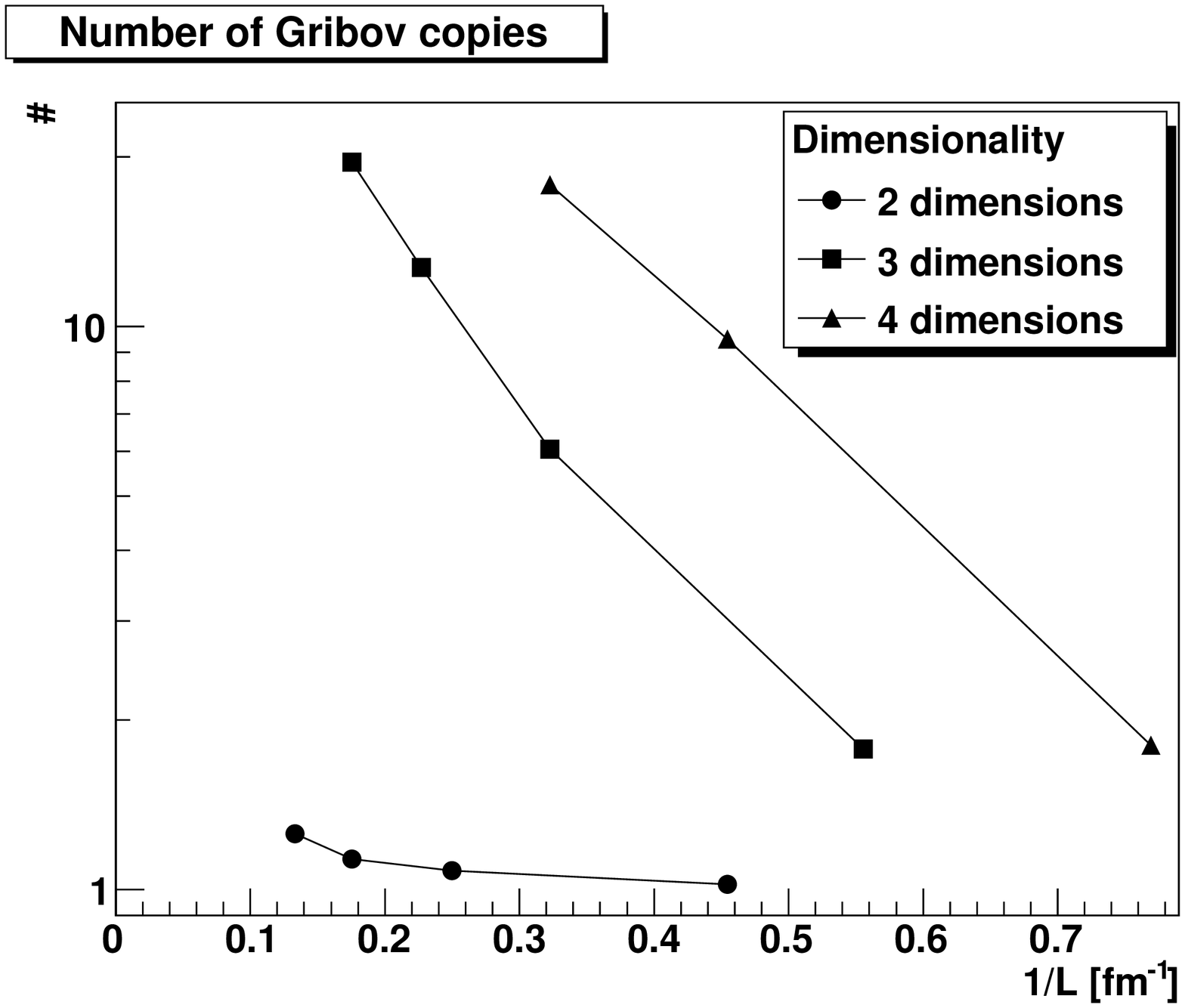}
\includegraphics[width=0.5\linewidth]{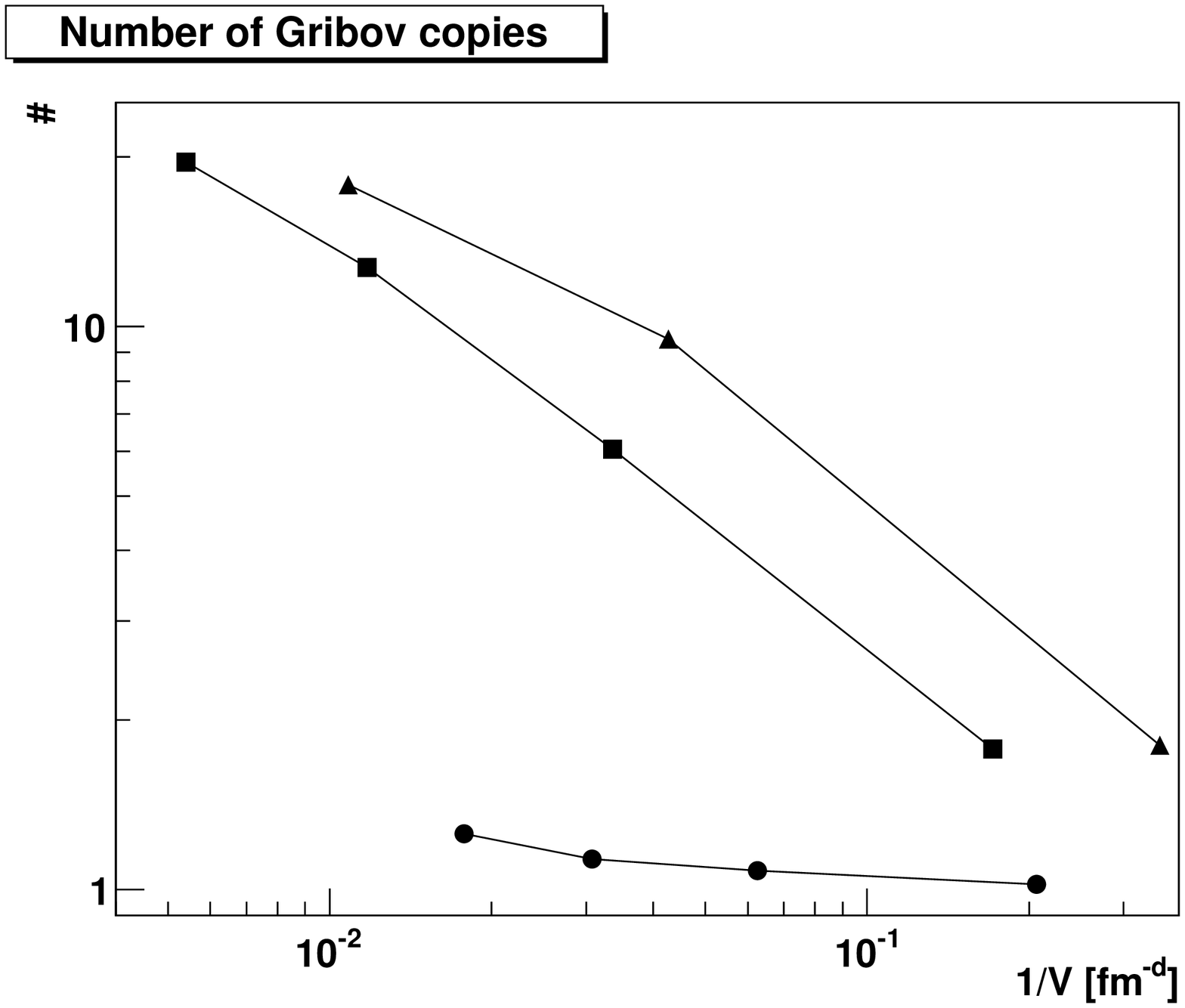}
\caption{\label{fig:gc}The average number of Gribov copies as a function of lattice extension (left panel) and volume (right panel) in two, three, and four dimensions at $a=0.22$ fm, distinguished by their value of $\tr D$. At finer $a$ at fixed physical volume the number of copies increases further \cite{maas}. All lines drawn to guide the eye. Lattice volumes $N^d$ are for $d=2$, 3, and 4 from the sets $\{10,18,26,34\}$, $\{8,14,20,26\}$, and $\{6,10,14\}$, respectively, throughout. A fit of type $A\exp(V/e^d)$ gives an $e$ of approximately 16, 4.4, and 2.9 fm in 2, 3, and 4 dimensions, respectively.}
\end{figure}

\begin{figure}
\includegraphics[width=\linewidth]{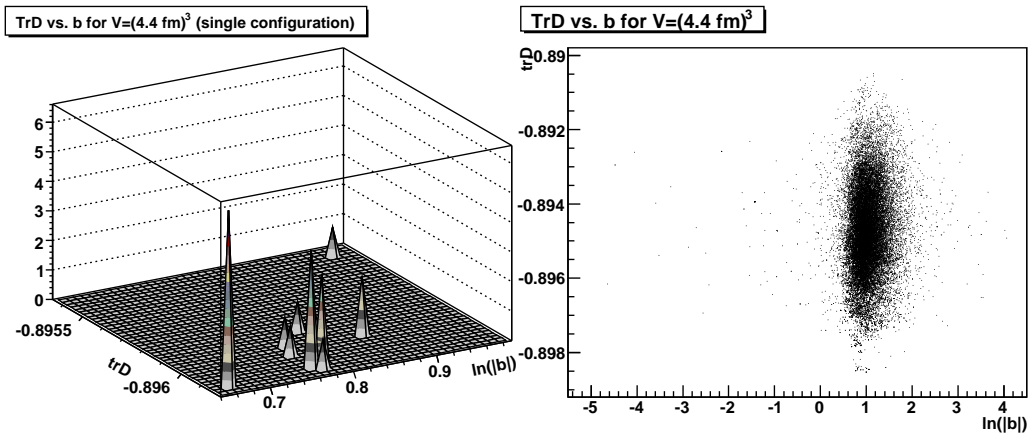}
\caption{\label{fig:gd}The distribution of $\tr D$ compared to $b(280 $ MeV$,\infty)$ on a 20$^3$ lattice at $a=0.22$ fm. The left panel shows for an example configuration that for each distinct Gribov copy, identified by differing $\tr D$ values, also the $b$ values differ. 22 copies have been generated for this configuration, with 10 distinct Gribov copies found, one peak being hidden by foreground peaks. The right panel shows the distribution for 1450 configurations.}
\end{figure}

Still, inside the first Gribov region there are many Gribov copies \cite{vanBaal:1997gu}. On a finite lattice\footnote{For details of the lattice calculations see \cite{Maas:2008ri,Cucchieri:2006tf} and for the generation of different Gribov copies \cite{Cucchieri:1997ns}.}, it appears that Gribov copies can be distinguished by their difference in the value of the trace of the gluon propagator \pref{trd} \cite{vanBaal:1997gu,Cucchieri:1997ns}. This will be assumed henceforth. The number of Gribov copies rises quickly, see figure \ref{fig:gc}, and thus there is a dense, though not necessarily connected, set of them in the infinite-volume limit\footnote{In two dimensions for the volumes investigated in \cite{Maas:2007uv} only a negligible amount of Gribov copies are present \cite{Maas:2008ri}. In these volumes a so-called scaling-type behavior \cite{Fischer:2008uz} prevails.}. Here, it is investigated whether other distinctions between Gribov copies exist. Motivated by the continuum studies \cite{Boucaud:2008ky,Fischer:2008uz} the ghost propagator was evaluated. It has been checked whether the Gribov copies can be characterized by the renormalization-group invariant quantity\footnote{This definition is prone to violations of rotational symmetry. For simplicity here all ghost momenta are evaluated along the $x$-axis, but a more robust method is required \cite{maas}.}
\be
b(p,P)=\frac{G(p)}{G(P)}=\frac{p^2D_G(p)}{P^2D_G(P)}\label{bdef}
\ee
\no for fixed $p\neq P$ and $G(p)=-p^2 D_G(p)$ is the ghost dressing function. If both momenta in the definition \pref{bdef} are large and comparable, $b$ will be of order one up to sub-leading corrections. From a practical point of view it is therefore better to use $p$ small and $P$ large. Here, $p$ will be taken to be the smallest accessible momentum, zero in the infinite-volume limit, and $P$ the renormalization scale $\mu$ taken in the perturbative domain, which can be chosen infinite in dimensions smaller than 4. In that case, $G(P)=1$ trivially, and the normalization can be dropped altogether. Note that the choice directly influences the severity of finite-volume artifacts, as they are stronger the smaller $p$ is chosen. Other choices will be studied elsewhere \cite{maas}.

Indeed, Gribov copies with differing $\tr D$ are found to have also differing values of $b$, as can be seen for an example configuration in the left panel of figure \ref{fig:gd}. The quantity $b$ measures in a sense the proximity to the Gribov horizon to the extent that it is dominated by the lowest eigenvalue and eigenstate of the Faddeev-Popov operator. It is found that $\tr D$ and $b$ turn out to be almost uncorrelated, see figure \ref{fig:gd} right panel. This would imply that there is little relation between the values of $b$ and the so-called fundamental modular region of minimal $\tr D$ values. However, it cannot be excluded that this is a lattice (discretization or finite volume) artifact.

\subsection{Constructing a gauge}

Since $b$ appears to be different for each Gribov copy, at least on the volumes investigated here, this would be sufficient to construct a non-perturbative extension of the Landau gauge, called Landau-$B$ gauges hereafter. It is defined by requiring that the function $b(0,\mu)$ ($b(\min p,\mu)$ on the lattice) should have a prescribed value of $B$ on the average. This is implemented by the following prescription: Take from each residual gauge orbit the copy with $b$ closest to $B$, and average over these copies to obtain the gauge-fixed quantities. In case of exactly the same difference for two copies, select one randomly to ensure the correct value of B on the average. By construction, the ghost propagator will then satisfy $b=B$ on the average, if the value of $B$ lies within the range of possible values for $b$. Otherwise a value as close as possible to $B$ will be obtained.

This prescription is valid for any volume, and specializes the perturbative Landau gauge \pref{landau} if more than one possible value of $b$ exists. If each Gribov copy of a given orbit has a unique value of $b$, this actually resolves the Gribov ambiguity completely. If not, this implies that further additional conditions can be imposed to lift the remaining degeneracies, specializing the Landau-$B$ gauges defined here further. If, like here, this is not done and there are degeneracies, just a random representative among the remaining degenerate Gribov copies is chosen. This is the same as done in minimal Landau gauge \cite{Cucchieri:2006tf}, where not even different values of $b$ are distinguished.

Note that both, Landau-$B$ gauges and minimal Landau gauge, are not affected by any Gribov ambiguity anymore: Without specifying how to resolve potential further degeneracies they yield a result which gives an expectation value over degenerate Gribov copies. In a thermodynamic setting this would be the equilibrium result, i.\ e.\ the most likely one, for correlation functions, provided the implemented algorithm faithfully represents the distribution of Gribov copies. In particular, no Gribov problem, i.\ e.\ the necessity to know all Gribov copies, arises in general. This problem is present, e.\ g., in the absolute Landau gauge or if the most extreme values of $B$ are desired.

Here 'equilibrium' is not indicating the existence of a preferred value: In the sense of stochastic quantization \cite{Zwanziger:2003cf}, no preferred Gribov copy on the orbit exists, any choice is equally admissible. Specifying a unique copy would therefore imply a maximum non-equilibrium situation, which is nonetheless as valid a choice as the equilibrium value of the minimal Landau gauge.

\subsection{Connecting to functional equations}

If this indeed identifies a gauge also in the infinite-volume and continuum limit this implies that the value of the ghost propagator distinguishes different treatments of Gribov copies. By analogy imposing conditions on the ghost propagator as boundary conditions on the functional equations, as detailed in \cite{Fischer:2008uz}, would then be equivalent to treat Gribov copies in such calculations. This would give an interpretation for the fact that the value of $B$ cannot be determined self-consistently inside functional equations \cite{Fischer:2008uz}: $B$ has the meaning of a second non-perturbative gauge parameter, besides the conventional gauge parameter $\xi$ to distinguish covariant gauges. This prescription is not unique, for alternatives see \cite{Maas:2008ri,Fischer:2008uz}. Note that choosing $B$ to be the average value among Gribov copies then provides a prescription how to implement the minimal Landau gauge in functional methods. If further degeneracies would exist, this interpretation would imply that with improved truncations compared to \cite{Fischer:2008uz} further undetermined boundary conditions must appear in the functional equations to permit resolving these degeneracies.

\section{Results for the correlation functions}

\begin{figure}
\includegraphics[width=0.5\linewidth]{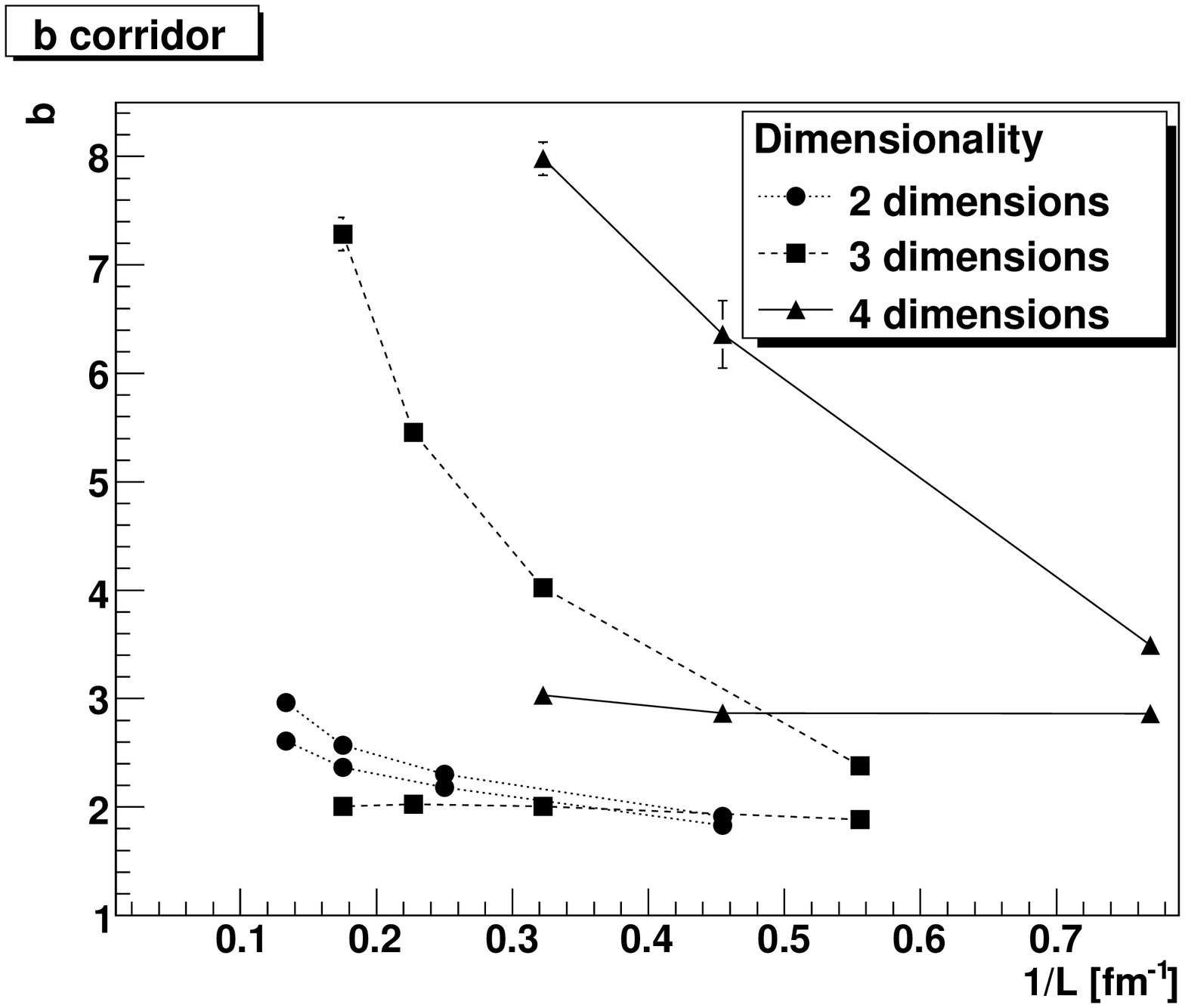}
\includegraphics[width=0.5\linewidth]{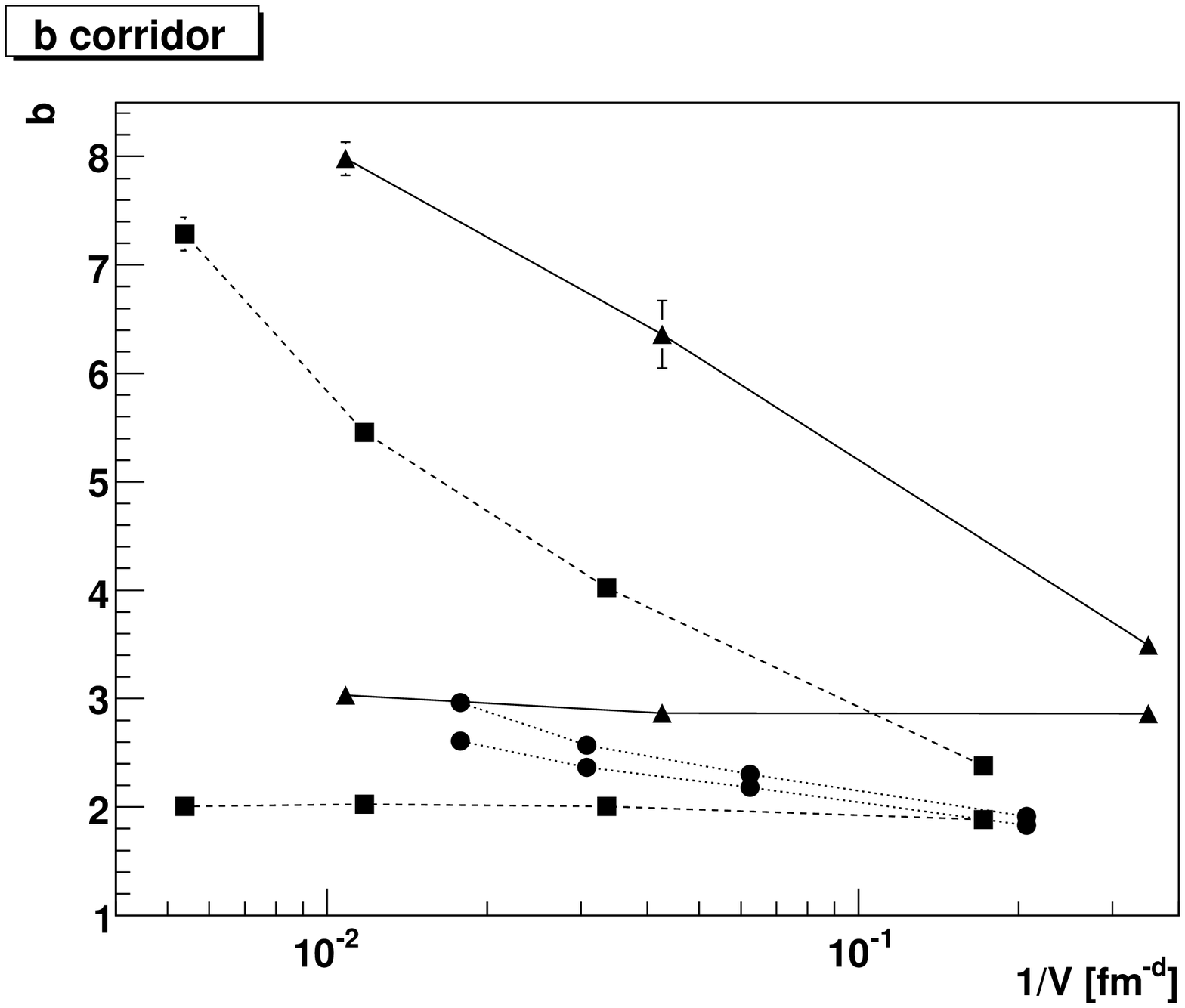}
\caption{\label{fig:gcorr}The gauge corridor for different lattice extensions (left panel) and volumes (right panel). For each volume $<\min b(\min p,\mu)>$ and $<\max b(\min p,\mu)>$ are shown, where $\min p$ is the smallest non-zero momentum on each lattice, and $\mu$ is set to $\infty$ for two and three dimensions, yielding $G(\mu)=1$, and to $\mu=2$ GeV in four dimensions.}
\end{figure}

The value of $B$ can only be chosen in a gauge corridor of values which are assumed on the residual gauge orbit. This corridor\footnote{Inherently, this method to identify Gribov copies \cite{Cucchieri:1997ns} cannot guarantee to find all copies. Hence, the boundaries of the corridor represent upper bounds for the lower boundary and lower bounds for the upper boundary. Finding the exact ones is a Gribov problem.} is shown in figure \ref{fig:gcorr}. Of course, if the corridor should decay into several bands with forbidden ranges in between, which is not observed here, choosing $B$ in between bands yields possibly ambiguities. In three and four dimensions the lower bound of the corridor is almost independent of volume, and by construction bounded from below. The upper bound is strongly increasing with volume for those investigated here. In two dimensions, due to the near-lack of Gribov copies at these volumes, the effect is rather small. Whether the upper boundary diverges in the infinite-volume limit remains to be investigated \cite{maas}. Without explicit exploration it cannot be excluded that the corridor closes again at much larger volumes, corresponding to total degeneracies of Gribov copies with respect to $b$.

\begin{figure}
\includegraphics[width=0.5\linewidth]{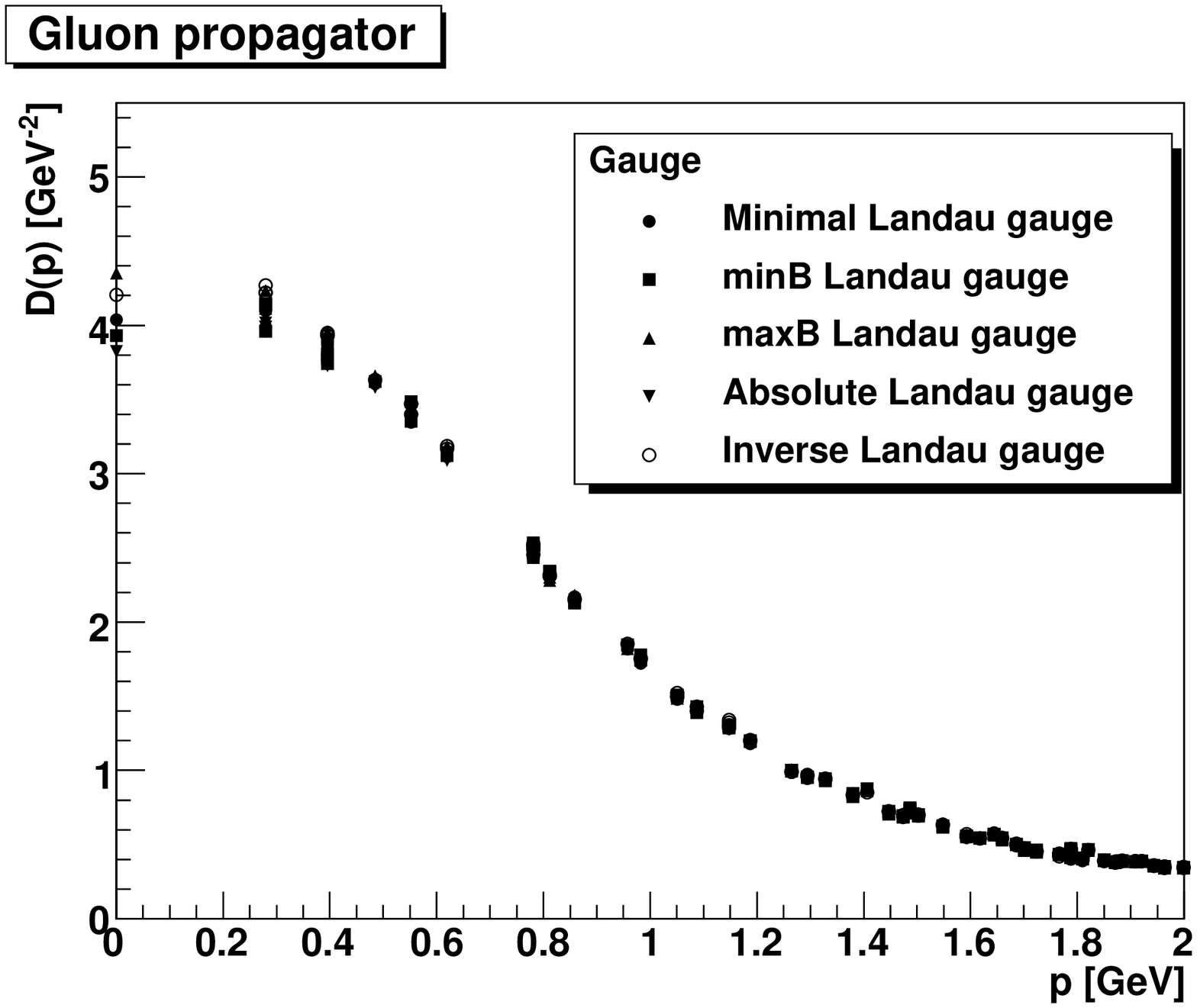}\includegraphics[width=0.5\linewidth]{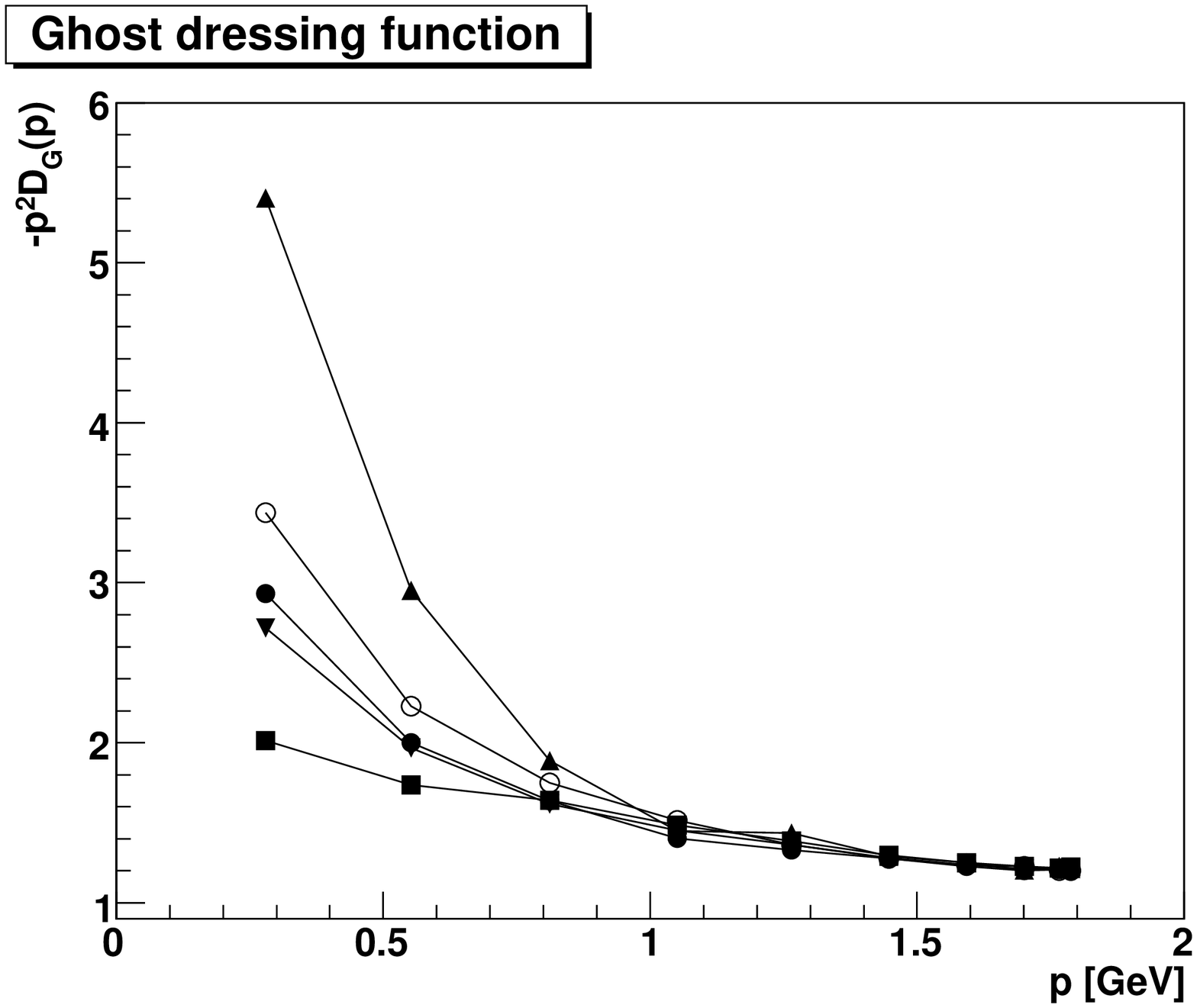}
\caption{\label{fig:corr}The gluon propagator (left panel) and the ghost dressing function (right panel) for various gauges from a 20$^3$ lattice at $a=0.22$ fm. The minimal Landau gauge is described in \cite{Cucchieri:2006tf}, and the absolute Landau gauge in \cite{Maas:2008ri}. The $\min B$ and $\max B$ gauges select the minimum and maximum possible value of $B$, respectively, and are described in the text. The inverse Landau gauge maximizes \pref{trd} over the set of all local minima on the residual gauge orbit \cite{Silva:2004bv}.}
\end{figure}

To compare extreme cases, the correlation functions for the cases with average, and minimum and maximum attainable value of $B$, corresponding to setting $B$ to zero and infinity, respectively, are determined. The results for the gluon propagator and the ghost dressing function in these three types of Landau-$B$ gauges, compared also to gauges based on \pref{trd}, are shown in figure \ref{fig:corr}. The ghost propagator differs very strongly at momenta below 1 GeV for the three Landau-$B$ gauges. In fact, the $\max B$-case is enormously enhanced compared to the result in minimal Landau gauge\footnote{Note that the bounds of \cite{Cucchieri:2008fc} are trivially fulfilled for any $B$-gauge, as they depend on the chosen $B$-gauge.}. The gluon propagator is, at least for this lattice size, almost indistinguishable for all gauges. This could have been anticipated due to the lack of correlation of the ghost and gluon propagator gauge parameters, shown in figure \ref{fig:gd}. Arguments exist \cite{Zwanziger:2003cf} that it, and any finite moments, may coincide for any fixing of the residual gauge degree of freedom in the infinite volume limit.

As the experience shows (see e.\ g.\ \cite{Cucchieri:2008fc}), rather large volumes will be necessary to be able to extrapolate to infinite volume. This will be addressed in the future \cite{maas}. However, there are in principle three qualitatively different outcomes, if no further degeneracies exist. First, the corridor is of finite width with finite boundaries. Then a family of decoupling-type correlation functions is obtained. Second, the upper boundary is infinite, and the ghost propagator corresponding to $1/B=0$ behaves as in the scaling case. The behavior of the gluon propagator is then not yet fixed, see the next subsection. Finally, the possibility exists that the upper bound is infinite, but the ghost propagator is not that of the scaling solution. If additional degeneracies exist, other possibilities are possible. E.\ g., the $b$-corridor can close to a single value, finite or infinite. In this case, only one $B$-gauge survives, with a unique ghost and gluon propagator.

\subsection{Relation to functional results}

Assume for the moment the correctness of the functional results yielding a one-parameter family of solutions with the scaling solution as an endpoint \cite{Fischer:2008uz}. Also an infrared finite gluon propagator is not at odds with this, as functional studies imply such a behavior at small volumes \cite{Fischer:2007pf}. Even if the gluon propagator would remain infrared finite in the infinite-volume limit (and possibly $B$-independent), this would not be at odds with this, since this (corresponding to a critical exponent \cite{Lerche:2002ep} of 1/2) is still compatible with a scaling-type behavior in case of mild angular variations of the ghost-gluon vertex \cite{Lerche:2002ep}. This is compatible with existing results \cite{Cucchieri:2006tf,Maas:2007uv,Cucchieri:2008qm}, and is expected from functional studies \cite{Schleifenbaum:2004id}. The role of the absolute Landau gauge is then not clear, as it so far appears unrelated to the $B$-gauges. In particular, it could well be that the interpretation given in \cite{Maas:2008ri} is wrong, and the absolute Landau gauge is not connected to a scaling behavior. This requires more investigations \cite{maas}.

\section{Summary}

Thus, using lattice gauge theory an one-parameter family of correlation functions at finite volume is found. These are distinguished by a second gauge parameter $B$ imposed on the ghost propagator. If there is no further freedom in choosing a Gribov copy\footnote{In the case that further degeneracies exist, there would be even more freedom to vary correlation functions by choice of Gribov copies.}, this would resolve the Gribov-Singer ambiguity completely. In addition, this would be achieved by conditions on the correlation functions only.

As a consequence, even the qualitative infrared behavior of correlation functions could be determined by a gauge choice, and any physics contained in there would be mixed with gauge contributions. A fuller analysis of this question will require more detailed studies \cite{maas}, and in particular much larger volumes. This applies also to the connection to a corresponding family of correlation functions obtained in functional studies \cite{Fischer:2008uz}, which could here only be established at the investigated volumes. If this could also be established for larger volumes and eventually in the infinite-volume limit, this would be extremely attractive. In particular, an infrared diverging ghost dressing function is a necessary condition for the Kugo-Ojima construction \cite{Kugo}. Another is a non-perturbatively well-defined BRST symmetry \cite{Fischer:2008uz}. Such a symmetry has been obtained \cite{Neuberger:1986xz}, but is based on an average over Gribov copies. Thus subtle cancellation between Gribov copies is required to bring this construction into contact with the one presented here.

If all of this would be correct in the end, this would be almost too good to be true: This would permit to use the gauge freedom to choose a gauge with useful properties. E.\ g., the explicit construction of the Hilbert space of QCD could be addressed by the use of the scaling (infinite $B$) Landau-$B$ gauge \cite{Fischer:2008uz} and the Kugo-Ojima construction \cite{Kugo}, which could provide one explanation of the confinement mechanism. Of course, also in the other Landau-$B$ gauges gluon \cite{Fischer:2008uz,Cucchieri:2004mf} and quark \cite{Braun:2007bx} confinement is obtained, though the explanation will necessarily be different.

{\bf Acknowledgments}

I am grateful to C.\ S.\ Fischer, H.\ Gies, J.\ Greensite, {\v S}.\ Olejn\'ik, J.\ M.\ Pawlowski and L.\ von Smekal for helpful remarks. This work was supported by the FWF under grant number M1099-N16. The ROOT framework \cite{Brun:1997pa} has been used in this project.


\end{document}